\documentclass[final]{svjour2}
\usepackage{graphicx}
\usepackage{rotating}
\usepackage{amssymb}
\usepackage{mathptmx}
\usepackage[numbers]{natbib}
\makeatletter
\journalname{Journal of Low Temperature Physics}

\bibpunct{}{}{,}{s}{}{,}

\begin{document}

\newcommand{\hdblarrow}{H\makebox[0.9ex][l]{$\downdownarrows$}-}
\title{Polarization  Effects in Superfluid $^4$He}

\author{V. P. Mineev}

\institute{Commissariat \`a l'Energie Atomique,
INAC/SPSMS, 38054 Grenoble, France
\\
Tel.:33-(0)4-38784409\\ Fax:33-(0)4-38785098\\
\email{vladimir.mineev@cea.fr}
}

\date{15.10.2010}

\maketitle


\begin{abstract}
A theory of thermoelectric phenomena in superfluid $^4He$ is developed. It is found an estimation of the dipole moment of helium atom arising due to electron shell deformation caused by pushing forces from the side of its surrounding atoms. 
The corresponding electric signal generated in a liquid consisting of electrically  neutral atoms  by the ordinary sound  waves is found extremely small. The second sound waves in superfluid $^4He$  generate the polarization of liquid  induced by the  relative accelerated motion of the superfluid and the normal component.   
The derived ratio of the amplitudes of temperature and electric polarization potential   was proved to be  practically temperature independent. Its magnitude is in reasonable correspondence with the experimental observations. The polarity of electric signal is determined by the sign of temperature gradient in accordance with the measurements. The problem of the roton excitations dipole moment is also discussed.

\end{abstract}

\section{Introduction}
The interaction of electromagnetic field  with superfluid $^4He$ is the subject of experimental studies performed in recent years by A. S. Rybalko and collaborators. There were discovered two groups of phenomena.
First, there was shown  that the second sound waves  as well  the liquid flow produced by the torsion oscillator and characterized by the relative non stationary motion of normal and superfluid component 
are accompanied by the  dynamic polarization of the liquid. \cite{Rybalko04,Rybalko05,Rybalko07,Rybalko10}   The inverse effect of second sound wave excitation by the periodic  electrical polarization of liquid  was also observed.  The experiments with the first sound show no electric  signal even with high powers applied to the sound emitter.

The second impressive finding was the discovery of resonant absorption of microwaves at the frequency, which corresponds to the roton minimum of the superfluid helium excitation spectrum. \cite{Rybalko2007} The inverse effect of generation of em radiation with roton frequency pumped  by heat flow in the superfluid helium has been detected. \cite{Ryb1,Ryb2} The following investigations have demonstrated the splitting of the roton resonance absorption line by the constant electric field.\cite{Rybalko2009}

The electric activity of neutral liquid consisting of helium atoms in ground state 
is quite astonishing property specific for the superfluid state of $^4He$. The observed effects are still  unexplained although there  were undertaken several attempts to interpret   the experiments with second sound (see the recent paper by Tomchenko \cite{Tom} and references therein) and the experiments with resonant em waves absorption at roton frequency \cite{Min}.

Here we propose  a theory of thermoelectric phenomena in superfluid $^4He$. It is shown that  
the electric signal generated in a liquid consisting of electrically  neutral atoms  by the ordinary sound  waves is extremely small. On the other hand, the second sound waves in superfluid $^4He$ can generate the polarization of liquid  accessible to observation. 
 It is caused by the  relative accelerated motion of the superfluid and the normal component.  
The relationship of amplitudes of temperature and electric potential oscillations is established. 

\section{Dipole moment induced by electronic shell deformation}

The helium atom in the ground state  do not possess a dipole moment. 
 The atom dipole moment  can arise  due to the electronic shell deformation.
Inside of liquid such a deformation is 
caused by the pushing force ${\bf f}$ acting on given atom  electron shell from the side of its surrounding atoms. In presence of an electric field ${\bf E}$ the 
total potential energy of external forces acting on helium atom is 
\begin{equation}
-({\bf f}-e{\bf E})({\bf r}_1+{\bf r}_2)-e{\bf E}{\bf r}_n,
\end{equation}
where ${\bf r}_1$ and ${\bf r}_2$
are the electron coordinates, ${\bf r}_n$ is the coordinate of nucleus and $e>0$ is elementary charge.
One can  pass to the coordinates of the electrons relative to nucleus $\mbox{\boldmath $\rho$}_1={\bf r}_1-{\bf r}_n$ and
$\mbox{\boldmath $\rho$}_2={\bf r}_2-{\bf r}_n$, and the coordinate of centre of gravity ${\bf R}=[m_e({\bf r}_1+{\bf r}_2)+
m_n{\bf r}_n]/(2m_e+m_n)$.  Performing this transformation we find 
that atomic Hamiltonian falls into the sum of two independent parts: the centre of gravity Hamiltonian and the Hamiltonian of two electrons consisting of the unperturbed part and the potential of perturbation determined by the external electric field and the force ${\bf f}$ acting on electrons 
\begin{equation}
\hat H_e=\hat H_0+(e{\bf E} -{\bf f})(\mbox{\boldmath $\rho$}_1+\mbox{\boldmath $\rho$}_2).
\end{equation}
The second term  here is written in usual approximation  that the electron mass is much smaller than the mass of nucleus $m_e\ll m_n$. 

The atom dipole moment is given by the average of $-e(\mbox{\boldmath $\rho$}_1+\mbox{\boldmath $\rho$}_2)$ over the ground state wave function given by
\begin{equation}
\Psi_0^\prime\approx\Psi_0-\frac{\langle\Psi_0|(e{\bf E} -{\bf f})(\mbox{\boldmath $\rho$}_1+\mbox{\boldmath $\rho$}_2)|\Psi_1\rangle}{E_1-E_0}\Psi_1,
\end{equation}
where $E_0,~E_1$ are the energies, and $\Psi_0, ~\Psi_1$ are the wave functions of the ground and the first excited states of helium atom in absence of perturbation.  Here we neglect by admixture of higher excited states.
The correction to the atom energy is
\begin{equation}
E_0^\prime-E_0\approx-\frac{r_{at}^2(e{\bf E} -{\bf f})^2}{E_1-E_0},
\end{equation}
where $r_{at}$  is the size of electron wave function of helium atom which is order of the hard core radius in the potential of interaction between two helium atoms. The linear in respect of electric field ${\bf E}$ term determines
the atom dipole moment  arising under the action of the force ${\bf f}$
\begin{equation}
{\bf d}=-\alpha{\bf f},~~~~~~\alpha\approx\frac{2 er_{at}^2}{E_1-E_0}.
\label{dipole}
\end{equation}
To estimate the numerical value of coefficient $\alpha$ we use $r_{at}\approx 2.7$ Angstrom,  $(E_1-E_0)\approx 20~eV$ and obtain $\alpha\approx 2.2\times10^{-14}$ CGSE units.

\section{First sound waves}

In the first sound waves each element of liquid moves with  acceleration  determined by the local value of the pressure gradient
\begin{equation}
{\bf a}=\frac{1}{\rho}\frac{\partial{\bf j}}{\partial t}=-\frac{\nabla P}{\rho}.
\end{equation}
Here, $\rho$ and ${\bf j}$ are  the liquid  denstity and density of current, 
and $P$ is pressure. It does not mean, however, that the each atom  electronic shell is deformed  under  the force $m{\bf a}$, where $m$ is the atom mass. Indeed,  during a period of the sound wave  each atom experience a lot of collisions with other atoms  such that the most of time it moves by inertia. Hence, the average force of the electronic shell deformation is weakened roughly speaking in $\omega\tau$ times
\begin{equation}
{\bf f}_1\approx\omega\tau~ m{\bf a}=-\frac{\omega\tau}{n}~\nabla P,
\end{equation}
where $n=\rho/m$ is the number of helium atoms in the unit volume, $\omega$ is the sound frequency and $\tau$ is the interatomic collision time. In liquid it   can be estimated as the  inverse Debye frequency $\tau\approx\omega_D^{-1}$. 

The electric field inside of liquid 
caused by the polarization of its atoms is
\begin{equation}
{\bf E}=-4\pi n{\bf d}=4\pi n\alpha ~{\bf f}_1\approx -4\pi \alpha~\omega\tau~\nabla P.
\end{equation}
One can  estimate the sound wave pressure amplitude which is sufficient to create an electric signal 
of measurable value. The sensibility to the voltage signal  $U_{min}$ in measurements done by Rybalko group  \cite{Rybalko04} was about 3 nanovolt. The  corresponding pressure amplitude 
\begin{equation}
\delta P=\frac{U_{min}}{4\pi \alpha\omega\tau}
\end{equation}
is about several Atmospheres at sound frequency   100 MHZ .   So, we see that the experimental registration of polarization generated by  waves of ordinary sound is quite difficult task. 

To describe the polarization effects generated by the second sound waves we shall discuss first the polarization  of atoms surrounding the single roton excitation.

\section{Roton dipole moment.}

 Couple years ago A.S.Rybalko and coauthors \cite{Rybalko2007} have revealed a resonant absorption of microwaves at the frequency corresponding to the roton minimum in the superfluid helium excitation spectrum. This finding was followed by the discovery of 
 splitting of the roton resonance absorption line by the constant electric field 
 \cite{Rybalko2009}. It is proved directly proportional to applied electric field $\pm dE$ as if a roton excitation has definite or "quantized" value of dipole moment $d\approx 10^{-22}$ CGSE units.

The explanation  of the experimental observations  has been proposed in the authors  paper \cite{Min}. There was pointed out that the roton excitation can not possess a stationary dipole moment, however, "a roton has an instantaneous dipole moment, such that at any given moment one can find it in the state either with positive or with negative dipole moment projection on its momentum direction". 
Following to Onsager\cite{Ons} and Feynman \cite{Feyn} hypothesis  a roton was treated \cite{Min} as the vortex ring of atomic size. An atom passing through the ring center, first acquires a momentum of the order of 
inverse interatomic distance $\hbar/a$ from the surrounding atoms and then losses momentum returning it to them.
The time dependent force corresponding to the periodic process of acceleration and deceleration of helium atoms passing through the vortex ring centre is estimated as
\begin{equation}
{\bf f}(t)\approx g(t)\frac{\hbar^2}{ma^3}\hat {\bf k}_r,
\end{equation}
where $\hat {\bf k}_r$ is the direction of roton momentum and $g(t)=\sum_{n\ge1}c_n\sin\omega t$ is a periodic function with period $\tau=2\pi/\omega\approx ma^2/\hbar$. Hence, according to relationship (\ref{dipole}) a roton  acquires the dipole moment
\begin{equation}
{\bf d}(t)=-\alpha g(t)\frac{\hbar^2}{ma^3}\hat {\bf k}_r.
\end{equation}
Although the magnitude of  dipole moment
\begin{equation}
d=\alpha f=0.7\times 10^{-22}g(t)~ed. CGSE 
\end{equation}
approximately coincides with the experimentally observed value \cite{Rybalko2009} the theory \cite{Min} 
cannot be recognized as satisfactory. The point is that the roton has to be stationary formation  whereas the proposed theory describes it as a non stationary object.

In the papers by L.Reatto et al\cite{Galli,Reatto} there was found the stationary variational many particle wave function of roton elementary excitation. It is characterized by the stationary inhomogeneous dipole like distribution of local atomic velocity ${\bf v}({\bf r})$ inside of roton wave packet with size of couple interatomic distances. 
This type of atomic motion can be also  thinkable as  caused by a field of local forces  ${\bf f}({\bf r})=m{\bf v}_i\nabla{\bf v}_i$ acting on atom  in point ${\bf r}$ inside of wave packet region.  Thus, each roton is accompanied by distribution of dipole moments of atoms ${\bf d}({\bf r})=-\alpha{\bf f}({\bf r})$ in surrounding liquid. The total dipole moment of a roton wave packet  is equal to zero 
\begin{equation}
\int{\bf d}({\bf r})d^3r=0. 
\end{equation}
This conclusion is certainly true for isolated roton which  has zero group velocity in infinite liquid.  Near the walls the roton wave packet is deformed and the roton acquires a dipole moment.  Similar effect one can expect in presence of normal liquid acceleration in respect of superfluid component  that  serves as 
an external field inducing  ordering of the directions and changing of the magnitudes  of dipole moments of atoms surrounding each roton. Hence, a relative superfluid-normal liquids  accelerated motion gives rise the roton dipole moment.   

This is apparently confirmed by the conservation laws   consideration \cite{Min}. Indeed, the photon momentum is many orders less than the roton momentum, hence, the creation of a single roton by means of single photon absorption is possible only if the excess of momentum is absorbed by the  liquid flow. It  also  makes possible the inverse process of the  photons generation induced by the roton annihilation decelerating  the relative superfluid - normal liquids flow. The quantum transitions in the superfluid liquid between the two states  with the same energy but different momenta was considered by G.E.Volovik \cite{Vol}. This process is accompaned by formation of semicircular quantized vortex near the wall of the container.

The submillimeter em waves - roton excitation-absorption line \cite{Ryb1} is much  narrow than the width of the roton absorption found in  inelastic neutron scattering experiments. An explanation of this as well as the reason why the observed roton 
dipole moment takes "quantized" 
values $ d~\approx~\pm10^{-22}$  CGSE units  are still absent.

\section{Thermoelectric waves}

In the second sound waves the local equilibrium is established by the fast  collisions in the gas of phonon-roton excitations.  They results in slow relative  accelerated motion of the superfluid and the normal component . Unlike the first sound waves the latter  is non-sensitive to the interatomic collisions and 
plays the role of an external field. 
We assume that an average dipole moment per atom induced by force related to
the relative accelerated motion of the superfluid and the normal components is
\begin{equation}
\langle {\bf d}\rangle=-\alpha{\bf f}_2=-\alpha m\frac{\partial ({\bf v_s}-{\bf v}_n)}{\partial t}=-m\frac{S}{\rho_n}\nabla T
\label{f2}
\end{equation}
Here, ${\bf v}_s$ and ${\bf v}_n$ are superfluid and normal liquid velocities, $\rho_n$ and $S$ are the 
normal component and the entropy densities and $T$ is temperature. The last equality in this formula is known equation of two fluid hydrodynamics\cite{Khal}.

In the second sound waves the electric field inside of liquid 
caused by the polarization of its atoms is
\begin{equation}
{\bf E}=-4\pi n\langle{\bf d}\rangle=4\pi \alpha n{\bf f}_2= 4\pi \alpha \rho\frac{S}{\rho_n}\nabla T 
\end{equation}
The direction of this field coincides with direction of the temperature gradient in correspondence with experimental observations
(induced negative charge moves to the hot wall )  \cite{Rybalko04}.
The ratio of the amplitudes of temperature to the electric potential is
\begin{equation}
\frac{\delta T}{\delta U}=\frac{{\rho_n}}{4\pi \alpha \rho{S}}
\end{equation}
The experiments \cite{Rybalko04,Rybalko05,Rybalko07,Rybalko10} have been performed in roton temperature region. Using the expression for the entropy and the normal component densities for the gas of rotons
\cite{Khal} one can rewrite this expression as
\begin{equation}
\frac{\delta T}{\delta U}=\frac{p_0^2}
{12\pi \alpha \rho
(\Delta+3T/2)}
\label{ratio}
\end{equation}
Here, $p_0$  is   roton momentum and  $\Delta$ is the gap in the roton spectrum. The only temperature dependent combination in this equation is $\Delta(T)+3T/2$. In temperature interval $1.4 - 1.8 K$ where the measurements were performed\cite{Rybalko04}  it changes from $10.8K$ to $11 K$. It means that in this temperature interval $\delta T/\delta U$ is practically  temperature independent that corresponds to the experimental observation\cite{Rybalko04}.  

To find the numerical value of the amplitudes ratio (\ref{ratio}) we take $p_0/\hbar=1.93$ Angstrom and $\rho=0.125g/cm^3$ and  obtain
\begin{equation}
\frac{\delta T}{\delta U}\approx 0.55\times 10^4~\frac{Kelvin}{Volt}
\end{equation}
This is about four times smaller than the experimentally found ratio \cite{Rybalko04}. The result is quite satisfactory in view of approximate estimation of $\alpha$ coefficient.

\section{Conclusion}

Based on simple idea of the atomic electron shell deformation
a  theory of thermoelectric phenomena in superfluid $^4He$ is developed.  The deformation arises due to the forces acting on the electron shell of given atom from the side of its nearest neighbours
in presence of accelerated motion of  liquid. In the first sound waves an atom   accelerated motion is result of fast chaotic  inter atomic collisions drastically decreasing the average local polarization of the liquid.
In the second sound waves  the local equilibrium is established by the fast processes of collisions in the gas of roton-phonon excitations.   The slow   relative accelerated motion of the superfluid and the normal component
plays the role of an external field non-sensitive to the interatomic collisions.
The force of deformation proportional to the normal - superfluid liquids relative acceleration causes  the appearance of atomic dipole moments  laying together in the macroscopic polarization.
The 
ratio of the amplitudes of temperature to the induced electric potential  in temperature interval $1.4 ~- ~1.8~ K$ proves   practically temperature independent that corresponds to the experimental finding. Its magnitude is  found in reasonable correspondence with the observations.

The previously found estimation \cite{Min} of "instantaneous" value of roton dipole moment has the  reasonable correspondence with experiments\cite{Rybalko2009}. However, the time average of this dipole moment is equal to zero as it should be for the roton considered as  stationary elementary excitation in superfluid $^4He$. Unlike to a roton at the rest  a roton accerated by the liquid flow acquires a dipole moment concentrated inside of its wave pocket having the size of the order several interatomic distances. According to experimental observation it seems that the roton dipole moment has fixed, or "quantized" values. The problems of  "quantized" values of roton dipole moment as well as of quite narrow em radiation roton absorption line  still remain to be solved.

The polarization effects in superfluid $^4He$ certainly deserve  of  further experimental studies, 
those hopefully stimulate theoretical investigations of microscopic mechanism of polarization phenomena.

\begin{acknowledgements}
I would like to thank A. S. Rybalko for the enlightening information and as well H. Godfrin who has attracted my attention to the L. Reatto papers.
\end{acknowledgements}

\pagebreak

\end{document}